\documentclass[epsfig,12pt]{article}
\usepackage{graphicx}
\usepackage{epsfig}
\usepackage{graphicx}
\usepackage{amssymb}
\usepackage{amsmath}
\usepackage{mathtools,physics,slashed}
\usepackage{hyperref}
\usepackage{float,subcaption}
\usepackage{tikz}
\usepackage[nottoc,notlof,notlot]{tocbibind} 
\usepackage[titles]{tocloft}

\allowdisplaybreaks
\usepackage{array}
\usepackage[margin=1in]{geometry}
\usepackage[numbers,sort&compress]{natbib}

\newcommand{\beq}{\begin{equation}}
\newcommand{\eeq}{\end{equation}}
\newcommand{\beqn}{\begin{eqnarray}}
\newcommand{\eeqn}{\end{eqnarray}}

\usepackage[textsize=scriptsize]{todonotes}

\def\d{\partial}
\def\pce{\psi\chi\eta}
\def\ZZ{\mathbb{Z}}
\def\RR{\mathbb{R}}
\def\cB{\mathcal{B}}

\def\cL{\mathcal{L}}
\def\cM{\mathcal{M}}
\def\cR{\mathcal{R}}

\begin{document}
\unitlength = 1mm

\begin{titlepage}

    \begin{flushright}
    FTPI-MINN-22-37, UMN-TH-4140/22\\
    \end{flushright}
    
    \vspace{2mm}
    
    \begin{center}
    {  \Large \bf  
    Consistency of \boldmath{$\chi$}SB in chiral Yang-Mills theory\\[1mm] 
    with adiabatic continuity
    }

    \vspace{5mm}
    
    {\large \bf   Chao-Hsiang Sheu$^{a}$ and Mikhail Shifman$^{a, b}$}
    \end {center}
    
    \begin{center}
    
        {\it  $^{a}$Department of Physics,
    University of Minnesota,
    Minneapolis, MN 55455}\\{\small and}\\
    {\it  $^{b}$William I. Fine Theoretical Physics Institute,
    University of Minnesota,
    Minneapolis, MN 55455}\\
    \end{center}
    
    \vspace{1cm}
    
    \begin{center}
    {\large\bf Abstract}
    \end{center}
    
        \vspace{5mm}
    
  We study the pattern of chiral symmetry breaking ($\chi$SB) in the $\psi\chi\eta$ model (with the chiral fermion sector containing $ \psi^{\{ij\}}$, $\chi_{[ij]}$, and $\eta_{i}^{A}$, see \cite{Armoni:2012xa}) on $\RR^3 \times S^{1}_{L}$ and derive  implications to $\mathbb{R}^4$ physics. Center-symmetric vacua are stabilized by a double-trace deformation. With the center symmetry maintained at small $L(S^1)\ll \Lambda^{-1}$, i.e. at weak coupling, no phase transitions are expected in passing to large $L(S^1)\gg \Lambda^{-1}$ (here $\Lambda$ is the dynamical Yang-Mills scale). Starting with the small $L$-imit, we find the leading-order nonperturbative corrections in the given theory. The instanton-monopole operators induce the adjoint chiral condensate $\langle \psi^{\{ij\}}\chi_{[jk]}\rangle \neq 0$  at weak coupling i.e. at  $L(S^1)\ll \Lambda^{-1}$. Then adiabatic continuity tells us that $\langle \psi^{\{ij\}}\chi_{[jk]}\rangle \neq 0$ exists on $\RR^4$, in full accord with the prediction \cite{Bolognesi:2017pek}. Simultaneously with $\langle \psi^{\{ij\}}\chi_{[jk]}\rangle \sim \Lambda^3\delta^i_k$ the SU($N_c$) gauge symmetry is spontaneously broken at strong coupling down to its maximal Abelian subgroup.

\end{titlepage}
\newpage

\section{Introduction}

Chiral Yang-Mills theory is an important element in understanding non-Abelain gauge dynamics. If at weak coupling (i.e., in the Stanadard Model) everything is transparent this cannot be said about the strong-coupling regime. In a special  ``hybrid" chiral model \cite{Armoni:2012xa}  (also known as the $\psi\chi\eta$ model) a pattern of the chiral symmetry breaking ($\chi$SB) was established \cite{Bolognesi:2017pek} on the basis of the 't Hooft anomaly considerations and some additional arguments. Two possible scenarios were revealed. In this paper, we invoke \"Unsal's adiabatic continuity for additional verification of the results reported in \cite{Bolognesi:2017pek}. The pattern of the chiral symmetry breaking in the $\psi\chi\eta$
model following from the adiabatic continuity perfectly coincides with one of the scenarios in \cite{Bolognesi:2017pek}.   We also discuss some other implications of  the adiabatic continuity
in the $\psi\chi\eta$
model.

The chiral Yang-Mills  theories were  studied at weak coupling and in various models (e.g., MAC) in the 1980s \cite{Raby:1979my,Geng:1986xh,Dimopoulos:1980hn} and more recently at strong coupling; see, e.g., \cite{Armoni:2012xa,Bolognesi:2017pek,Bolognesi:2019wfq,Bolognesi:2021jzs,Bolognesi:2022beq,Shifman:2008cx,Smith:2021vbf,Bolognesi:2021yni,Bolognesi:2020mpe,Bolognesi:2019fej,Anber:2021iip}.  Owing to the recent discoveries  of generalized global symmetries in quantum field theory \cite{Gaiotto:2014kfa,Gaiotto:2017yup}, and the corresponding  't Hooft matching, this topic has attracted renewed attention \cite{Bolognesi:2021yni,Smith:2021vbf,Poppitz:2009uq,Poppitz:2009tw}.

We will focus on one particular subclass of strongly coupled Yang-Mills theory with a special chiral fermions matter sector.

Two simplest cases (the so-called $\psi\eta$ and $\chi\eta$ model) were previously studied in \cite{Bolognesi:2021jzs,Smith:2021vbf,Poppitz:2009tw,Poppitz:2009uq,Bolognesi:2017pek,Bolognesi:2019wfq,Bolognesi:2021yni,Bolognesi:2020mpe} from various perspectives.
In particular, in \cite{Bolognesi:2017pek,Bolognesi:2019wfq}, under the assumption of a nonvanishing  condensate $\expval{\psi\eta}$ based on the traditional 't Hooft anomaly, matching  the color-flavor locking between the gauge SU$(N)$ and the global SU$(N + 4)$ was observed. The  $\chi$SB pattern then takes the form 
 $${\rm SU}(N)_{\rm c} \times {\rm SU}(N+4)_{\rm f} \times {\rm U}(1) \to {\rm SU}(N)_{\rm cf} \times {\rm U'}(1) \times {\rm SU}(4)_{\rm f} \,.$$
The infrared-bound  baryon states ($\psi^{\{ij\}}\eta_i^A\eta^B_j$) were constructed. 

Then, the authors of \cite{Bolognesi:2020mpe} argued, on the basis of a mixed 't Hooft anomaly, that a chirally symmetric vacuum is impossible. However, the authors of \cite{Smith:2021vbf} (based on recasting the center of the symmetry group)  raised objections concerning the result of Bolognesi {\em et al}. \cite{Bolognesi:2020mpe}.
In view of the ongoing debate on the vacua of   the  $\psi\eta$ and $\chi\eta$ models 
we will leave the simplest cases aside for the time being and limit ourselves to  the hybrid $\psi\chi\eta$ version in the framework of the adiabatic continuity approach.

\"{U}nsal suggested \cite{Unsal:2007jx,Unsal:2007vu} the adiabatic continuity method \cite{Unsal:2008ch,Shifman:2008cx,Shifman:2008ja,Poppitz:2009tw,Poppitz:2009uq} to study strongly coupled gauge theories. See \cite{Poppitz:2021cxe} for the recent review. 
The starting point of this method is as follows. 
Consider theories on the spacetime with one spatial dimension compactified, say, $\RR^3 \times S^{1}_L$. Unlike compactified  time dimension in thermalized theories \cite{Gross:1980br,Weiss:1980rj}, adiabatic continuity construction was argued to contain no  phase transition in passing from the small $L$ limit $L(S^1)\ll\Lambda ^{-1}$ to large $L$ -- approaching $\RR^4$ -- provided the vacuum of the model at small $L(S^1)$ is center symmetric. In this case the vanishing of the Polyakov line is preserved intact at both limits.
Then the model Abelianizes, all gauge bosons outside the Cartan subalgebra acquire masses
and the theory becomes weakly coupled, on the one hand, and still confining according to the Polyakov criterion. 
Perturbative calculations turn out to be reliable. Moreover, some nonperturbative features such as instanton monopoles and bions can be explicitly computed
in the quasiclassical approximation  and provide nontrivial results for four-dimensional physics upon continuation to large   $L(S^1) \gg \Lambda^{-1}$. In Sec.~\ref{sec:adiabatic}, we will show how the instanoton-monopole operators allow us to verify the $\chi$SB pattern found in \cite{Bolognesi:2017pek}.

In particular, in the $\pce$ model, three types of monopole operators show up --only one of which is a global gauge rotation singlet participating in the low-energy effective Lagrangian. It is such an $\cM_j$-type monopole that induces chiral condensates $\expval{\psi\chi}$. The dynamical Abelianization  takes place at large $L(S^1)$ and hence on  $\RR^4$ by the standard lore of adiabatic continuity. This verification perfectly matches the recent study of $\chi$SB in the $\pce$ model via the mixed 't Hooft anomaly \cite{Bolognesi:2022beq}. 

In the first part of the paper, the center-symmetry stabilization is achieved by virtue of a double-trace deformation \cite{Unsal:2008ch,Unsal:2007jx}.
In the second part, we replace the double-trace deformation by additional fermions (additional with regards to the primary chiral matter), which do the same job. We find a few exceptional  cases when the additional fermions can stabilize center symmetry. 
We note a potential disadvantage of implementing stabilization by virtue of additional adjoint fermions and appropriate  flavor-twisted boundary conditions for (chiral)  fermions in the fundamental representation.

The paper is organized as follows. In Sec.~\ref{sec:pcereview}, we briefly review  the $\pce$ model and some previously conjectured $\chi$SB  patterns. In Sec.~\ref{sec:adiabatic}, the technique of adiabatic continuation is applied to examine possible scenarios of $\chi$SB with the purpose of corroborating them.

In Sec. \ref{sec:adjstab}, the interplay between the stabilization strategy (double-trace deformation vs additional fermions) is discussed and the corresponding fermion condensates are found in two examples.
Our conclusions are summarized in Sec.~\ref{sec:summary}.

\section{The \boldmath{$\pce$} Model}\label{sec:pcereview}

In this section, we review some basic elements of the $\pce$ model playing the main role in the following discussion. The $\pce$ model refers to a chiral SU$(N_c)$ gauge theory with the fermion sector consisting of a pair of two-index fermions, one symmetric while the other antisymmetric in color indices, plus eight antifundamentals. Namely, 
\begin{align}
    \psi^{\{ij\}}, \qquad
    \chi_{[ij]}, \qand
    \eta_{i}^{A}
    \qfor 
    A = 1,2,...,8
\end{align}
where $\psi^{\{ij\}}$ is the symmetric fermion and $\chi_{[ij]}$ is antisymmetric.
The model is self-consistent and enjoys the symmetry structure \cite{Armoni:2012xa,Bolognesi:2017pek,Bolognesi:2019wfq,Bolognesi:2022beq}
\begin{align}
    G = \frac{
        {\rm SU}(N_c) \times {\rm SU}(8)_{f} \times {\rm U}(1) \times \tilde{\rm U}(1)
        }{
        \ZZ_{N_c} \times \ZZ_{8/N^{*}}
    }
    \label{2}
\end{align}
in which $N^{*}$ stands for the greatest common divisor of $N_c+2$ and $N_c-2$. The overlapped phase rotations of fermions between the centers and U(1) groups are quotiented out.
 
The $\psi\chi\eta$ model has three U(1) symmetries, with the generating currents
\beq
j^{\dot\alpha\alpha}_{(\psi )} = \bar \psi^{\dot\alpha}\psi^\alpha\,,\qquad
j^{\dot\alpha\alpha}_{(\chi )} = \bar \chi^{\dot\alpha}\chi^\alpha\,, \qquad
j^{\dot\alpha\alpha}_{(\eta )} = \bar \eta^{\dot\alpha}\eta^\alpha\,,
\label{4}
\eeq
(no summation over the flavor index $A$ in the definition of $j^{\dot\alpha\alpha}_{(\eta )}$).
Each of the above currents is anomalous,
\begin{align}
    \partial_{\alpha\dot\alpha} j^{\dot\alpha\alpha}
    &= \partial_\mu\,j^\mu 
    \nonumber\\
    &= \mqty(
        N+2\,,& \mbox{ for}\,\,\psi\\[2mm]
        N-2\,,& \mbox{ for}\,\,\chi\\[2mm]
        1\,,& \mbox{ for}\,\,\eta\\[2mm]
    )
    \times \frac{1}{32\pi^2} \, F_{\mu\nu}^a \tilde{F}^{\mu\nu\,a}\,.
\end{align}
The last expression allows us to establish two nonanomalous combinations of the currents denoted by U(1) and 
$\tilde{\rm U}(1)$  
in Eq. (\ref{2}),
For instance, 
\beqn
&&{\rm U}(1): \quad (N-2) j_{(\psi)} - (N+2)  j_{(\chi)}\,,\nonumber\\[2mm]
&&\tilde{\rm U}(1): \quad 2 j_{(\psi)} -  2 j_{(\chi)}- \sum_A j_{(\eta)}\,.
\label{5p}
\eeqn
Equation (\ref{5p}) corresponds to the following
charge assignment of fermions:
\begin{align}\label{6p}
        {\rm U}(1):&\qquad Q_{\psi}=\frac{N_c-2}{N^{*}} \,,\quad Q_{\chi}=-\frac{N_c+2}{N^{*}} \,,\quad Q_{\eta}=0,
        \nonumber\\[1mm]
        \tilde{\rm U}(1):&\qquad \tilde{Q}_{\psi}=2 \,,\quad \tilde{Q}_{\chi}=-2 \,,\quad \tilde{Q}_{\eta}= -1 \,.
\end{align}

The phase rotation of fermions under the centers of SU$(N_c)$ and SU(8) together with the U(1)s takes the form
\begin{align}
    \psi &\to e^{2(N_c-2)\pi i\alpha/N^*}e^{4\pi i\beta}e^{4\pi ki/N_c}\, \psi \,,
    \notag\\[1mm] 
    \chi &\to e^{-2(N_c+2)\pi i\alpha/N^*}e^{-4\pi i\beta}e^{-4\pi ki/N_c} \,\chi\,,
    \\[1mm]
    \eta &\to  e^{-2\pi i\beta}e^{-2\pi ki/N_c}e^{\pi mi/4} \,\eta \,,
    \nonumber
    \label{four}
\end{align}
in which $\alpha,\beta \in (0,1)$, $k \in \ZZ \pmod{N_c}$, and $m \in \ZZ \pmod{8}$. Here, $\alpha$ and $\beta$ stand for the transformations under ${\rm U}(1)$ and $\tilde{\rm U}(1)$, respectively, while the integers $k,m$ parametrize the center transformation of ${\rm SU}(N_c)$ and ${\rm SU}(8)$ groups.
Solutions for $\alpha,\beta\neq 0$ exist, which leave (\ref{four}) intact; they are   parametrized by $\ZZ_{N_c} \times \ZZ_{8/N^{*}}$. See a more detailed discussion  in \cite{Bolognesi:2019wfq,Bolognesi:2022beq}.

As first noted in \cite{Bolognesi:2017pek} and further analyzed in \cite{Bolognesi:2019wfq,Bolognesi:2022beq}, the $\pce$ model {\em a priori} can have two different $\chi$SB  scenarios in the infrared regime. First, let us suppose both $\expval{\psi\eta}$ and $\expval{\psi\chi}$ develop nonzero vacuum expectation values (VEVs). The former condensate then implies  that the color and flavor indices are entangled in such a way  that the condensate is invariant under the diagonal SU(8) transformation. This is the so-called color-flavor locking. As a consequence, if $N_c \geq 12$, the chiral symmetry breaks as follows:
\begin{align}
    {\rm SU}(N_c) \times {\rm SU}(8)_{\rm f} \times {\rm U}(1) \times \tilde{\rm U}(1)
    \to 
    {\rm SU}(8)_{\rm cf} \times {\rm U}(1)^{N_c-p+1} \times {\rm SU}(p-8)_{\rm c} \,
\end{align}
where we ignore the center part. At $p=12$, the theory saturates the SU(8)$^3$ 't Hooft anomaly.
On the other hand, if only $\psi\chi$ has a nonvanishing VEV (i.e., $\expval{\psi\eta}=0$), 
we arrive at a Higgs/confinement phase.

Once the adjoint condensate $\expval{\psi^{ij}\chi_{jk}}$ develops, the pattern of $\chi$SB takes the form 
\begin{align}\label{eq:dymab}
    {\rm SU}(N_c) \times {\rm SU}(8)_{f} \times {\rm U}(1) \times \tilde{\rm U}(1) 
    \to 
    \prod_{l=1}^{N_c-1} {\rm U}(1)_{l} \times {\rm SU}(8)_f \times \tilde{\rm U}(1) \,.
\end{align}
The fundamental fermions  stay massless and weakly coupled with the massless dual photons in the infrared regime \cite{Bolognesi:2017pek}.

\section{Adiabatic Continuity Analysis}\label{sec:adiabatic}
In the following discussion, we will show that the above scenario is compatible with the prediction from the adiabatic continuity of the deformed $\pce$ model on $\RR^3 \times S^{1}_{L}$.
To proceed, let us consider the $\pce$ model with a double-trace deformation \cite{Shifman:2008cx,Shifman:2008ja,Unsal:2008ch}, say, 
\begin{align}\label{eq:defmaction}
    S =  \frac{1}{g^2} \int_{\RR^3 \times S^1_{L}} \dd^{4}{x} \, \left\{ \frac{1}{2}\Tr F^2 + \sum_{\cR} i\overline{\Psi}^{\cR}\slashed{D}\Psi^{\cR} \right\}
\end{align}
in which $\cR$ runs over the relevant irreducible representations of chiral fermions in the theory.
Along $S^{1}_L$, all fermions (as well as bosons) obey the periodic boundary condition.
$S_{\rm double\,\,trace}$ here is fine-tuned in such a way that the effective potential is stabilized\footnote{
    The effective potential $V_{\psi+\chi} \sim \sum_p (\tr_{F}\Omega^p)^2$ develops a local minimum at the center-symmetric point, and the double-trace term is only used to stabilize  $V_{\eta}$ generated by fundamental fermions as in \cite{Shifman:2008cx}.
\label{fn1}} at the center-symmetric point \cite{Argyres:2012ka}
\begin{align}
    \left\{ La_{j} \right\}_{j=1}^{N_c} = 
    \left\{ \frac{N_c-1}{N_c}\pi,~ \frac{N_c-3}{N_c}\pi,~ \cdots,~ -\frac{N_c-1}{N_c}\pi \right\} \,.
\end{align}
In the small-$L$ regime, the gauge symmetry $SU(N_c)$ is broken down to $[U(1)]^{N_c-1}$ at the center-symmetric point and the off-diagonal gauge bosons as well as fermions acquire three-dimensional masses.
In particular, the mass terms of two-index fermions $\psi^{ij}$ and $\chi_{ij}$ are 
\begin{align}
    m_{ij} = \frac{2\pi}{LN_c} (N_c+1-i-j)
\end{align}
implying $\psi$ and $\chi$ remain massless in perturbation theory if $i+j = N_c+1$. For later convenience, let us denote 
\begin{align}
    \psi^{j} := \psi^{\{j,N_c+1-j\}}
    \qand
    \chi_{j} := \chi_{[j,N_c+1-j]} \,.
\end{align}

Before directly getting into the chiral symmetry breaking in transition from the small radius to the four-dimensional limit, we first consider the model at small $L(S^1)$, at weak coupling.
It is straightforward to obtain the perturbation theory  by integrating the modes along the $S^{1}_{L}$ direction. On the other hand, we emphasize that the building blocks to describe no-perturbative dynamics of the $\pce$ model on $\RR^3 \times S^1_{L}$ are the monopole-instanton operators. In SU$(N_c)$ gauge theory, we deal with  $(N_c-1)$ BPS monopoles and one KK monopole \cite{Lee:1997vp,Lee:1998bb,Kraan:1998pm}.
The fermion zero mode distribution can be found from the index theorem \cite{Nye:2000eg,Poppitz:2008hr}. Namely, 
\begin{align}\label{eq:monopoles}
    \cM_{j} =& e^{-S_0+i\vb*{\alpha}_j\vb*{\sigma}} 
    (\psi^{j}+\psi^{j+1})(\chi_j + \chi_{j+1}),
    \notag\\[2mm]
    \cM_{N_c-1} =& e^{-S_0+i\vb*{\alpha}_{N_c-1}\vb*{\sigma}} (\psi^{1}+\psi^{2})^2 \eta^{A_1}\cdots\eta^{A_8} \epsilon_{A_1A_2\cdots A_8}, \nonumber\\[2mm] 
    \cM_{N_c} =& e^{-S_0+i\vb*{\alpha}_{N_c}\vb*{\sigma}} (\psi)^2
\end{align} 
for $j =1,2,..., N_c-2$. Note that $\vb*{\sigma}$ is the dual photon field in three dimensions and $\{\vb*{\alpha}_1,\vb*{\alpha}_2,...,\vb*{\alpha}_{N_c}\}$ are affine roots of the $su(N_c)$ algebra. In Eq.~\eqref{eq:monopoles}, we will keep only massless fermions for the zero modes as the leading-order contribution.
To see which of these monopole operators generate the relevant condensates, first recall that under the global $[{\rm U}(1)]^{N_c-1}$ gauge rotations, 
\begin{align}
    \psi^{\{ab\}} &\to e^{i(\vb*{H}_{aa}+\vb*{H}_{bb})\vb*{\xi}}\psi^{\{ab\}},
    \nonumber\\
    \chi_{[ab]} &\to e^{-i(\vb*{H}_{aa}+\vb*{H}_{bb})\vb*{\xi}}\chi_{[ab]},
    \\
    \eta_{a} &\to e^{-i\vb*{H}_{aa}\vb*{\xi}}\eta_{a}
    \nonumber
\end{align}
where $\vb*{H}$ is the elements of the Cartan subalgebra of $su(N_c)$.
Then, to construct a global-rotation invariant monopole operator, we have to integrate out all U$(1)$ phases, 
\begin{multline}
    e^{-S_0+i\vb*{\alpha}_j\vb*{\sigma}} 
    \cdot\int \,  
    \Biggl(
        \psi^j\chi_j
        + e^{i[\vb*{H}_{j+1,j+1}+\vb*{H}_{N_c-j,N_c-j} - (j \to j-1)]\vb*{\xi}}\psi^{j+1}\chi_j
        + (j \leftrightarrow j+1)
    \Biggr)
    \dd{\vb*{\xi}}
\end{multline}
which results in
\begin{align}\label{eq:monopolej}
    \cM_j = e^{-S_0 + i\vb*{\alpha}_j\vb*{\sigma}}\left( \psi^j\chi_j+\psi^{j+1}\chi_{j+1} \right).
\end{align}
All other operators vanish.
This is  different from the observations in \cite{Poppitz:2009tw,Poppitz:2009uq,Shifman:2008cx}.
In our model, a certain kind of monopole operator \eqref{eq:monopolej} dressed with chiral fermions forms a global gauge rotation singlet, while in the aforementioned publications, monopole operators with chiral zero modes do not survive and only act as a building block for higher-order operators.
In fact, this is not so surprising because $\psi^{ik}\chi_{kj}$ transforms as an adjoint boson, and we saw in many examples \cite{Argyres:2012ka,Anber:2017pak,Shifman:2008ja} that the instanton-monopole operators with the adjoint zero modes do play a role in the nonperturbative dynamics of the theory.

The existence of the $\cM_j$-type monopole operators then implies  the low-energy (large-distance) effective Lagrangian of the form 
\begin{align}\label{eq:dualLag}
    \cL_{\rm dual} = \frac{g_3^2}{32\pi^2}(\d_{\mu}\vb*{\sigma})^2 + \overline{\Psi}i\slashed{\d}\Psi
    + \left( \sum_j e^{-S_0 + i\vb*{\alpha}_j\vb*{\sigma}}\left( \psi^j\chi_j+\psi^{j+1}\chi_{j+1} \right)
    + \mbox{h.c.} \right) + \cdots
\end{align}
where the former two terms come from the perturbation theory while the dots stand for the higher-order nonperturbative contributions such as bions and instantons.

\vspace{1mm} 

Now, let us probe the pattern of $\chi$SB by taking into account the nonperturbative contributions mentioned above. In light of \eqref{eq:dualLag}, there exists a nonvanishing chiral condensate 
\begin{align}
    \expval{\psi^{ik}\chi_{kj}}
    \sim \expval{\psi^{l}\chi_{l}} \delta^{i}_{j}
    = c_{j}\Lambda^3e^{4i\alpha/N^{*}}\delta^{i}_{j}
    \label{19p}
\end{align}
with 
\begin{align*}
    \sum_{j=1}^{N_c} c_j = 0
\end{align*}
where $\Lambda$ is the strong scale. This is exactly the same as what was argued in \cite{Bolognesi:2017pek,Bolognesi:2022beq} and therefore dynamical Abelianization \eqref{eq:dymab} ensues.
Note that due to the shift symmetry, 
\begin{align}\label{eq:shiftsymm}
    &\vb*{\alpha}_j\vb*{\sigma} \to \vb*{\alpha}_j\vb*{\sigma} + \frac{4\alpha}{N^*} \,, \qquad
    \nonumber\\[1mm]
    &\vb*{\alpha}_{N_c-1}\vb*{\sigma} \to \vb*{\alpha}_{N_c-1}\vb*{\sigma} - \frac{2(N_c-2)\alpha}{N^*} - 4\beta \,, 
    \\[1mm]
    &\vb*{\alpha}_{N_c}\vb*{\sigma} \to \vb*{\alpha}_{N_c}\vb*{\sigma} - \frac{2(N_c-2)\alpha}{N^*} + 4\beta \,,
    \nonumber
\end{align}
the bion excitations can show up  in the spectrum and generate the dual photon masses. 
To be more precise, the magnetic bions are neutral under the shift symmetry and, according to Eqs.~\eqref{eq:monopoles} and \eqref{eq:shiftsymm}, the bion operators can be explicitly written in the form 
\begin{align}
    \cB_{ij} \sim e^{-2S_0}e^{i(\vb*{\alpha}_i-\vb*{\alpha}_j)\vb*{\sigma}}
    \label{18}
\end{align}
where $j$ can be $i+1$, $N_c-i$, $N_c-1-i$, or $N_c+1-i$. Note that $i,j$ are both smaller than $N_c-1$ since it is the pairing among $\cM_1,...,\cM_{N_c-2}$ as implied by the shift symmetry.
Thus, $\sigma_{1},...,\sigma_{N_c-2}$ become massive while $\sigma_{N_c-1}$ remains massless.
Those massive gauge bosons then result in the theory being confining by virtue of the Polyakov mechanism \cite{Polyakov:1976fu}.

We can also imagine a situation that the gauge symmetry $[{\rm U}(1)]^{N_c-1}$ is spontaneously broken. That is, we lift the constraints on the invariance of the instanton monopoles under the global gauge rotations. In this case, all instanton-monopole vertices in Eq.~\eqref{eq:monopoles}  participate in nonperturbative dynamics. The additional condensate is then 
\begin{align}
    \expval{(\psi^1+\psi^2)^2\eta^{A_1}\cdots\eta^{A_8} \varepsilon_{A_1A_2\cdots A_8}} \neq 0
\end{align}
coming from $\cM_{N_c-1}$.
Note that because the massless fermions ($\psi^1$ and $\psi^{N_c}$) do not couple to $\cM_{N_c}$, it cannot be included in the leading contribution in the sense that massive fermions decouple in the weak-coupling region.
Lastly, in the presence of the $\cM_j$ instanton monopoles within this framework, the suggested symmetry breaking pattern is identical to \eqref{eq:dymab} by the same argument. This is one of the scenarios first advocated in \cite{Bolognesi:2017pek}.

\section{Center Stabilization with Adjoint Fermions}
\label{sec:adjstab}

In this section, we will abandon the idea of the double-trace deformation and discuss stabilization problem in some models with chiral fermions in mixed representations --including adjoint fermions --and how these mixed ensembles  affect the prediction for 
$\chi$SB based on  adiabatic continuity.
Adding adjoint fermions is intended to stabilize center-symmetric vacua in the effective potential. As has been mentioned in footnote \ref{fn1}, there is no need to stabilize the  $\psi$-$\chi$ pair.

What could possibly help in this situation is to introduce the (flavor) twisted-boundary condition (FTBC) \cite{Cherman:2017tey,Kouno:2013mma,Misumi:2014raa}.\footnote{For recent developments on the implication of $\chi$SB  in QCD with fundamental or adjoint fermions through  introduction of  twisted boundary conditions see, \cite{Unsal:2021xay,Kanazawa:2019tnf}.}
While  in Sec. \ref{sec:adiabatic}, all fermions obey the periodic boundary condition
\begin{align}\label{eq:pbc}
    \psi(x^\mu,x^3+L) = \psi(x^\mu,x^3) \,,
\end{align}
now we switch on a flavor twist  for fundamentals to construct an  additional flavor-dependent phase.\footnote{One may want to consider the most general FTBC, $\eta^{A}(x_3+L) = U^{AB}\eta_B(x_3)$ with $U^{AB} \in SU(N_f)$. Yet, the twist matrix can always be diagonalized by making a flavor rotation.} Instead of  Eq. \eqref{eq:pbc}, now we require
\begin{align}
    \eta_{k}^A(x^\mu,x^3+L) = \pm\omega^A \eta_{k}^A(x^\mu,x^3)
    \qq{with}
    \omega^n = 1
    \,.
\end{align}
This then alters the fundamental fermion contribution to the effective potential
\begin{align}
    V_{\rm eff} \sim \sum_{p=1}^{\infty} \frac{1}{p^4} \tr_F\Omega^p
    ~\to~ 
    \sum_{p=1}^{\infty} \frac{1}{n^3p^4} \tr_F\Omega^{np}
\end{align}
so that more models fall into the (local) minimum at the center-symmetric point as extra adjoint fermions are added.
In the following we provide two examples with FTBC fundamental fermions and point out the shortcoming of this boundary condition in determining the schemes of $\chi$SB.

\begin{figure*}[t] 
        \centering 
        \begin{subfigure}[b]{.32\textwidth}
            \centering
            \includegraphics[width=.7\linewidth]{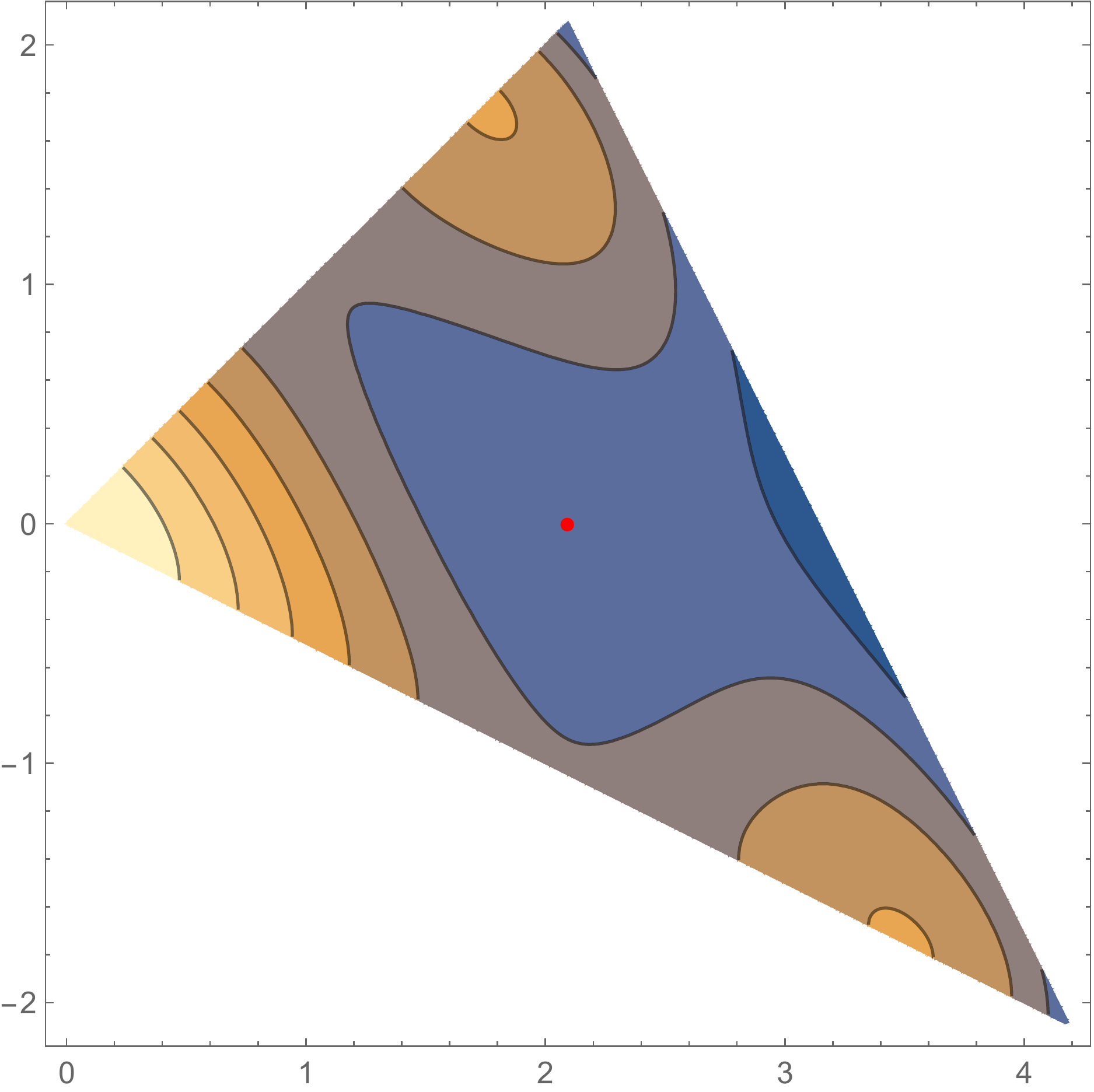}
            \caption{One extra adj. fermion}
        \end{subfigure}
        \hfill 
        \begin{subfigure}[b]{.32\textwidth}
            \centering
            \includegraphics[width=.7\linewidth]{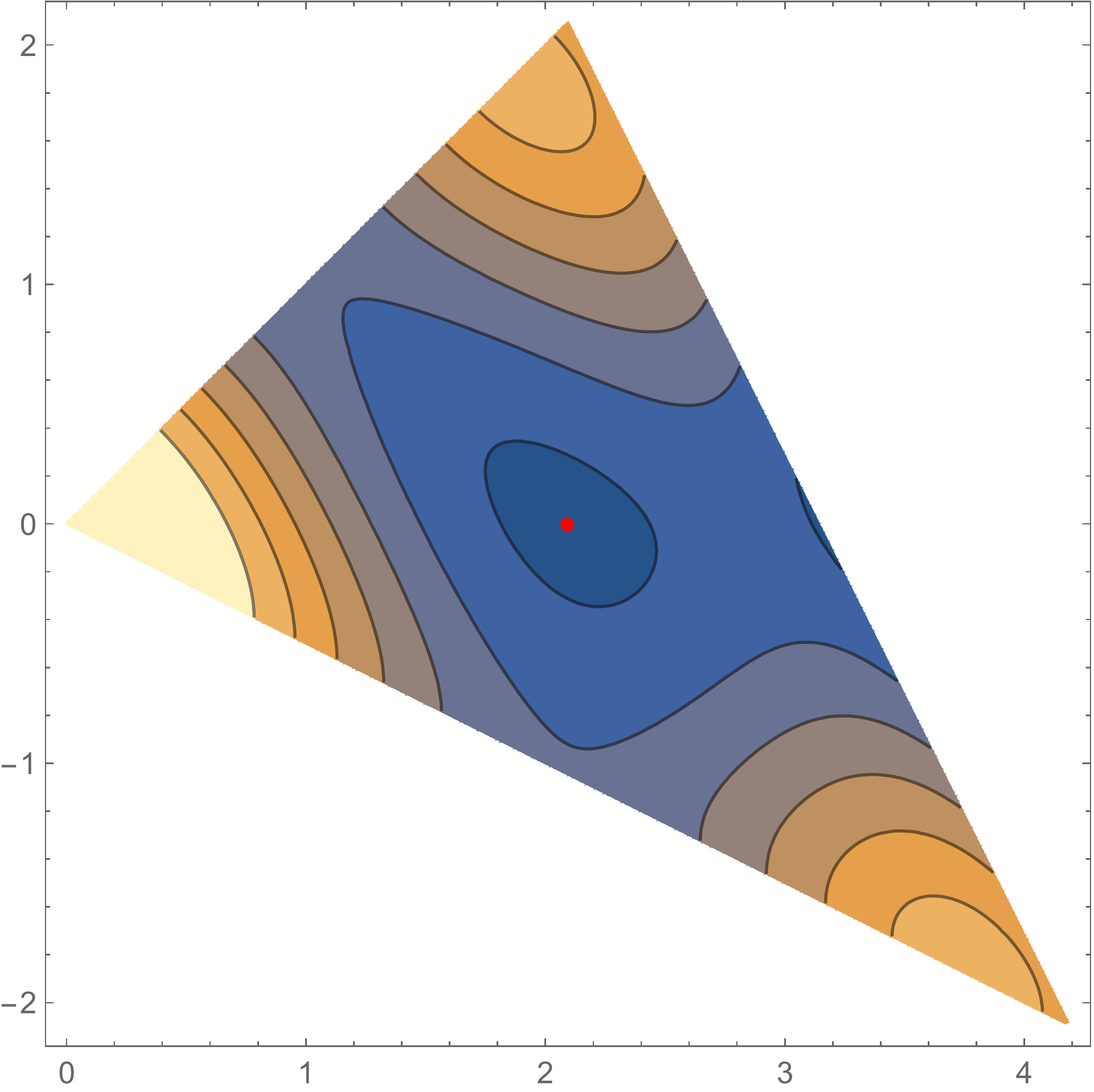}
            \caption{Two extra adj. fermions}
        \end{subfigure}
        \hfill 
        \begin{subfigure}[b]{.32\textwidth}
            \centering
            \includegraphics[width=.7\linewidth]{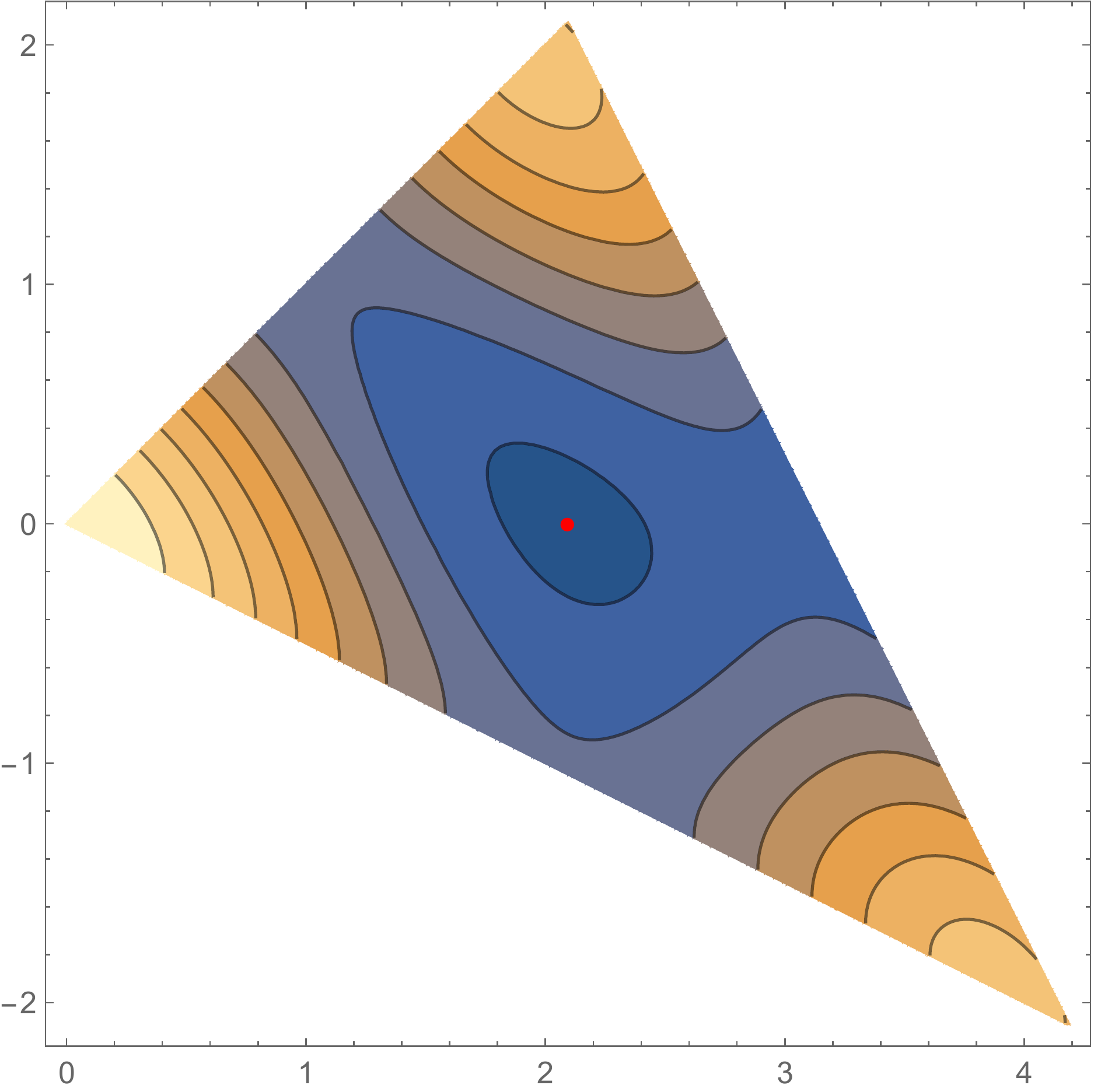}
            \caption{Three extra adj. fermions}
        \end{subfigure}
        \caption{\small The effective potentials of one, two, and three adjoint fermions added in the Weyl chamber. The horizontal and vertical axes represent two components of the background holonomy, respectively. The red dot illustrates the center-symmetric point.\label{fig:adjstable}}
\end{figure*}

First, let us consider  SU(3) Yang-Mills theory with the following fermion sector: $\psi^{ij}$, $\chi^{ij}$, and $\eta_{i}^{A}$ for $A=1,2,...,6$ where the fundamentals $\eta^{A}_k$ satisfy the boundary condition 
\begin{align}
    \eta_{k}^A(x^\mu,x^3+L) = e^{\frac{2\pi Ai}{3}} \eta_{k}^A(x^\mu,x^3)
    \,.
\end{align}
The effective potential with the varying number of the adjoint fermions is demonstrated in Fig. \ref{fig:adjstable}. The center symmetry is achieved  at least at the local minimum and as long as the number of adjoint fermions is greater than one it is in fact a global minimum.
Next, the instanton-monopole operators of the theory under consideration  take the form 
\begin{align}
    &\cM_1 = e^{-S_0}e^{i\vb*{\alpha}_{1}\cdot\vb*{\sigma}}
    \chi\psi\eta^{3}\eta^{6} 
    \,, \quad
    \cM_2 = e^{-S_0}e^{i\vb*{\alpha}_{2}\cdot\vb*{\sigma}}
    (\psi)^2\eta^{1}\eta^{4}
    \,, \quad
    \nonumber\\[1mm]
    &\cM_3 = e^{-S_0}e^{i\vb*{\alpha}_{3}\cdot\vb*{\sigma}}
    (\psi)^2\eta^{2}\eta^{5}\,.
\end{align}
Four-fermion zero mode vertices  appear for each monopole. The corresponding lowest-order condensate then should be a four-fermion composite so that it cannot provide further information as to the next-to-leading-order contribution because the minimal condensates argued in \cite{Bolognesi:2019wfq} consist of only two fermions. 

Another class of center-stabilized anomaly-free theories is SU$(N_c)$ gauge theories with one $\psi^{\{ij\}}$, a number of two-index antisymmetric $\tilde{\chi}^{[ij]}$ and $\chi_{[ij]}$ fields, and pairs of fundamental fermions satisfying periodic and antiperiodic boundary conditions. There is only a finite number of combinations for this kind of model with stabilization at the center-symmetric point, namely $N_c = 4,\,6,\,8,\,10,\,14,\,16,\, 20,\, 24,\, 36,\, 52$. The idea is to cancel the $\chi$ contribution driving the effective potential away from the center-symmetric point by pairs of periodic and antiperiodic fermions, namely,
\begin{align}
    -\frac{k-1}{2}(\tr_F\Omega^{2p})
        +\frac{n_f/2}{2^3} (\tr_F\Omega^{2p}) = 0
\end{align}
where $k$ is the total number of $\tilde{\chi}^{ij}$ and $\chi_{ij}$.
The distribution of the fermion zero modes on the monopole operators can be sketched as follows:
\begin{align}
    \cM_{j} &= e^{-S_0}e^{i\vb*{\alpha}_j\vb*{\sigma}} \psi\tilde{\chi}^{k_1}\chi^{k_2}
    \,, \quad
    \notag\\[1mm]
    \cM_{\frac{N_c}{2}} &= e^{-S_0}e^{i\vb*{\alpha}_{N_c/2}\vb*{\sigma}} \psi^2\eta_1\eta_2\cdots\eta_{n_f}
    \,, \quad
    \\[1mm]
    \cM_{N_c} &= e^{-S_0}e^{i\vb*{\alpha}_{N_c}\vb*{\sigma}}
    \psi^2\,.
    \notag
\end{align}
In the same manner as in the beginning of this section, the related condensates contain multifermion composites and turn out to be irrelevant in verifying the pattern of $\chi$SB.

In the above examples, the theories we have identified all have an exact center-symmetric vacuum. 
In fact, the condition that the effective potential has to stabilize at the center-symmetric point can be further relaxed as long as the Abelianization of the vacua (of a theory) persists. 
This then opens up more classes of theories whose chiral symmetry breaking can be verified via the adiabatic continuity.  At the moment, however, a systematic search is hardly possible in an analytic (or even a numerical) way because of the vagueness of the condition.

Summarizing, imposing the flavor-twisted boundary conditions allows a number of theories to be stabilized at the center-symmetric point. This circumstance makes possible the method of adiabatic continuity. However, the downside is  forcing the fundamental fermion zero modes to be evenly distributed among the monopole operators. The emerging condensates  then consist of a larger number of fermions than needed for the $\psi\chi $ condensate
(in the leading approximation). Therefore, the FTBC-based line of reasoning in the present form is useless
 for corroborating the pattern of the 
$\chi$SB which was argued for in Refs.~\cite{Bolognesi:2017pek,Bolognesi:2019wfq}.

\section{Comments and conclusions}\label{sec:summary}

In this work, we start from the UV symmetry group in the $\pce$ model and consider possible ways  of $\chi$SB on $\RR^4$. After a brief introduction, we discuss dynamics, especially nonperturbative aspects, in the $\pce$ model on $\RR^3 \times S^1_{L}$ at small $L$. We limit ourselves to the leading approximation (instanton monopoles). The center-symmetric vacuum is guaranteed by a double-trace deformation. The dynamical Abelianization is achieved. Then the corresponding instanton-monopole operators are found to induce the chiral condensate $\expval{\psi\chi}$. Adiabatic continuity propagates this result to $\RR^4$. This is a successful and desired part of our work. With $\langle \psi^{\{ij\}}\chi_{[jk]}\rangle \sim \Lambda^3\delta^i_k$ the SU($N_c$) gauge symmetry is spontaneously broken down  to its maximal torus $[{\rm U}(1)]^{N_c-1}$. This is our target --verification of the prediction \cite{Bolognesi:2017pek,Bolognesi:2019wfq} in Sec. \ref{sec:adiabatic}.

The strategy of abandoning   the double-trace deformation in favor of adding extra adjoint fermions and flavor twisting of boundary conditions for the fundamental fields $\eta$ for stabilization at small $L$ in essence failed.  Although stabilization can be achieved in a limited number of cases, no useful information can be obtained in this way about the condensate of interest.

In conclusion it is worth adding a comment about the planar equivalence between ${\mathcal N}=1$ Yang-Mills theory (without matter) and the $\psi\chi\eta$ model. The latter model was designed \cite{Armoni:2012xa} in the context of the concept of planar equivalence in the common sectors which, in the case at hand, covers all correlators of purely gluon operators. 

At small $L(S^1)$, the $\psi\chi\eta$ model is characterized by Polyakov confinement
due to bions [see Eq. (\ref{18})]. If the adiabatic continuity is correct we expect confinement at 
large $L(S^1)$ [on $\RR^4$ in the limit $L(S^1)\to\infty$]. However, because of the fact that $\langle \psi^{\{ij\}}\chi_{[jk]}\rangle \sim \Lambda^3\delta^i_k$,
dynamically this confinement is somewhat different from that in ${\mathcal N}=1$ Yang-Mills theory on $\RR^4$. One can call it the {\em Higgs/confinement phase}. The vacuum structure is also different: While in  ${\mathcal N}=1$ Yang-Mills theory we have $N$ vacua marked by the gluino order parameter,
$\langle \lambda\lambda\rangle \sim \Lambda^3 \exp \left(\frac{i 2\pi k}{N}
\right)$,  in the $\psi\chi\eta$ model the order parameter (\ref{19p}) continuously rotates under the action of the anomaly-free current
$\frac{1}{N^*}\left[(N-2) j_{(\psi)} - (N+2)  j_{(\chi)}\right]$; see (\ref{5p}) and (\ref{6p}). This implies a continuous 
 U(1) vacuum manifold which in turn implies the existence of a massless particle.

\section*{Acknowledgments}

We are very grateful to Mohamed Anber, Aleksey Cherman and Erich Poppitz for valuable discussions.
This work is supported in part by DOE grant DE-SC0011842.

\end{document}